\documentstyle[prl,aps,twocolumn]{revtex} 

\input{psfig.sty}

\begin{document}

\draft

\title {Strong enhancement of extremely energetic proton 
production in central heavy ion collisions at intermediate energy}

\author { P. Sapienza,$^1$ R. Coniglione,$^1$ M.~Colonna,$^1$   
E.~Migneco,$^{1,2}$ C.~Agodi,$^1$ R.~Alba,$^1$ G.~Bellia,$^{1,2}$ 
A.~Del~Zoppo,$^1$ P.~Finocchiaro,$^1$ V.~Greco,$^{1,2}$ 
K.~Loukachine,$^1$ C.~Maiolino,$^1$
P.~Piattelli,$^1$ D.~Santonocito,$^1$ P.G.~Ventura,$^1$ 
Y.~Blumenfeld,$^3$ M.~Bruno,$^4$ N.~Colonna,$^5$ M.~D'Agostino,$^4$ 
L.~Fabbietti,$^6$ M.L.~Fiandri,$^4$ F.~Gramegna,$^7$ I.~Iori,$^{6,8}$
G.V.~Margagliotti,$^9$ P.F.~Mastinu,$^7$ P.M.~Milazzo,$^9$ A.~Moroni,$^8$ 
R.~Rui,$^9$ J.A.~Scarpaci,$^3$ and G.~Vannini$^4$ }

\address { (1) INFN - Laboratorio Nazionale del Sud,
            Via S. Sofia 44, I-95123 Catania (ITALY)}
\address { (2) Dipartimento di Fisica dell'Universit\`a di Catania (ITALY)}
\address { (3) Institut de Physique Nucl\'eaire, IN2P3-CNRS-F-91406 Orsay, 
France}
\address { (4) INFN and Dipartimento di Fisica dell'Universit\`a di Bologna (ITALY)}
\address { (5) INFN - Sezione di Bari, Bari (ITALY)}
\address { (6) Dipartimento di Fisica dell'Universit\`a di Milano (ITALY)}
\address { (7) INFN - Laboratori Nazionali di Legnaro, Legnaro (ITALY)}
\address { (8) INFN - Sezione di Milano, Milano (ITALY)}
\address { (9) INFN - Sezione di Trieste, Trieste (ITALY)}

\date{\today}

\maketitle

\begin {abstract}
The energetic proton emission has been investigated as a function of the 
reaction centrality for the system $^{58}$Ni + $^{58}$Ni at 30A MeV. 
Extremely energetic protons ($E_p^{NN}\geq 130 MeV$) 
were measured and their multiplicity is found to increase
almost quadratically with the number of participant nucleons thus indicating
the onset of a mechanism beyond one and two-body dynamics. 
\end {abstract}

\pacs{PACS numbers: 25.70.-z, 25.75.Dw}

Heavy ion collisions at intermediate energy allow to investigate the
properties of
nuclear matter far from stability. Dynamical calculations show that, 
in the early non
equilibrated stage of the reaction, high temperatures and densities 
are reached. Since heavy ion reactions at intermediate 
energy are described in terms of
mean field and two body collisions, experimental observables 
sensitive to basic ingredients of the models such as the nucleon-nucleon 
(NN) cross section in matter and the mean field potential are needed to 
probe the nuclear dynamics and the equation of state of nuclear 
matter (EOS). In particular,
particles such as subthreshold mesons or energetic photons and 
nucleons are expected to provide information on the nuclear dynamics at the 
pre-equilibrium stage \cite{CAS90,GEL99} and reference therin.
Experimentally the presence of a pre-equilibrium component in the light 
particle emission has been observed in the energy spectra \cite{FUC94,WAD89}.
The impact parameter dependence of the pre-equilibrium protons, investigated 
for several systems, shows
that the average energetic proton multiplicity increases almost linearly 
with the number of participant nucleons \cite{ALB94}.
These features support the hypothesis of
pre-equilibrium proton emission due to incoherent quasi-free NN collisions 
\cite{FUC94}. 
Moreover, experimental evidences, 
such as the observed  $\gamma$-proton anti-correlation \cite{SAP94} and the 
energetic proton angular distributions reported in reference \cite{CON00},
provide information on the space-time origin of the energetic protons 
indicating that a relevant fraction is emitted from first chance NN 
collisions in the interaction zone. 
These evidences show that the energetic protons are emitted in the first 
stage of the reaction according to expectations \cite{CAS90} 
and therefore are good candidates
to probe the pre-equilibrium phase. 
On the other hand, the observation of extremely energetic nucleons or deep
subthreshold particles over a broad range of incident energy addresses 
the question of which mechanisms
could enable to concentrate a relevant fraction of the 
available energy in the production of a single energetic or massive particle 
\cite{MAR99}. In fact, the emission of particles with energy or 
mass much larger than that provided by the coupling of the 
nucleon Fermi motion with the relative motion of the colliding nuclei is a 
challenging aspect of heavy ion collisions both experimentally 
and theoretically, due to the very low production rates and the lack of 
information on the production mechanism. 

In this letter we present results concerning the emission of 
protons with energy extending up to almost 20$\%$ of the total available energy
in the reaction $^{58}$Ni + $^{58}$Ni at 30A MeV. Since these energies largely
exceed the maximum energy expected in first chance NN collisions due to the 
coupling of the relative motion with a sharp nucleon Fermi momentum 
distribution (kinematical limit),
this investigation can also provide a clue for
the comprehension of the deep subthreshold particle emission in this 
energy domain. 
In particular, for the first time, the proton multiplicity as a function
of the impact parameter 
was measured for extremely energetic protons ($E_p^{NN} \geq 130$ MeV).
A strong non linear dependence on the number of participant 
nucleons is observed thus providing 
important information on the production mechanism.
A detailed comparison with a microscopic transport model 
was also performed aiming to 
extract information on the dynamics at pre-equilibrium and the nature 
of energetic proton emission.

The experiment was performed at Laboratori Nazionali del Sud
with the MEDEA \cite{MIG92} and MULTICS \cite{IOR93} apparatus.
A $^{58}$Ni beam at 30A MeV delivered by the Tandem and Superconductive
Cyclotron (CS) acceleration system bombarded a $^{58}$Ni target 
2 mg/cm$^2$ thick.   
MEDEA consists of a ball made of 180 BaF$_2$ detectors placed at 
22 cm from the target which covers the polar angles from 30$^o$ to 170$^o$. 
The BaF$_2$ permits to detect and identify LCP 
($E_p \leq 300$ MeV) and photons up to $E_\gamma \approx 200$ MeV.  
The time of flight and pulse shape discrimination
analysis allows to clearly identify photons, protons, deuterons and tritons,
alpha particles \cite{MIG92}.
The response of BaF$_2$ crystals to LCP has been investigated using
monoenergetic particle beams and a 
calibration procedure based on the $\gamma$ calibration with $\gamma$ sources
and cosmic rays has been established as described in ref. \cite{ADZ93} .
The MULTICS array is made of 55 telescopes covering the angular 
range $3^o \leq \theta_{lab} \leq 28^o$. Each telescope consists of an
Ionization Chamber, a Silicon detector and a CsI crystal, and allows the 
identification of charged particles up to $Z=83$.
The threshold for charge identification was about 1.5A MeV \cite{IOR93}.
The total geometric acceptance was greater than 90$\%$ of 4$\pi$.

Energetic protons were detected in coincidence with photons,
light charged particles ($Z=1,2$) (LCP) and intermediate and heavy 
fragments on an event by event basis, thus allowing a rather complete 
description of the reaction dynamics as well as an estimate of impact parameter
which represents a crucial problem in this energy domain.  
Due to the very low cross section expected for extremely energetic protons, 
high statistics spectra are needed.
To increase the fraction of events containing energetic protons, the main
trigger required the presence of at least one BaF$_2$ signal above a threshold 
level corresponding to proton energy of about 30 MeV.
Moreover the coincidence with MULTICS reduced the 
cosmic ray contamination to a negligible level.
All the MEDEA detector with $\theta \geq 75^o$ and a few
detectors of the forward rings took part in this trigger. 
Events corresponding to a minimum bias 
trigger, defined by the OR between MEDEA and MULTICS, 
were also scaled down and registered.    
Altogether approximately $4 \times 10^8$ events have been collected 
and analysed.

According to the standard three moving source analysis,
the high energy proton emission at large polar angles can be
described by a source emitting with velocity close to the half beam 
velocity and a high inverse slope parameter.
A selection of the energetic protons emitted from this intermediate velocity
source is possible by applying
kinematical constraints
($E_p^{\rm lab} \geq 40$ MeV, $\theta_{\rm lab} \geq 42^o$)
\cite{SAP98}. The experimental proton spectra, transformed in the NN frame 
($v/c=0.127$)
are reported in fig.~1a (full symbols) together with the intermediate source
component of the fit (solid lines) for inclusive data.
The inverse slope parameter deduced from the maxwellian fit with a volume 
emission is in good agreement with the systematics ($T \approx 11$ MeV) 
\cite{FUC94}.
Very energetic protons are observed in the spectra, with energy well above 
the kinematical limit expected in the hypothesis of first chance NN 
collisions and sharp Fermi momentum distribution 
$(v_{max}=v_{F}+0.5v_{\rm beam})$
(arrows of fig.~1). 

\begin{figure}[ht]
\vspace {0.0cm}
\psfig{figure=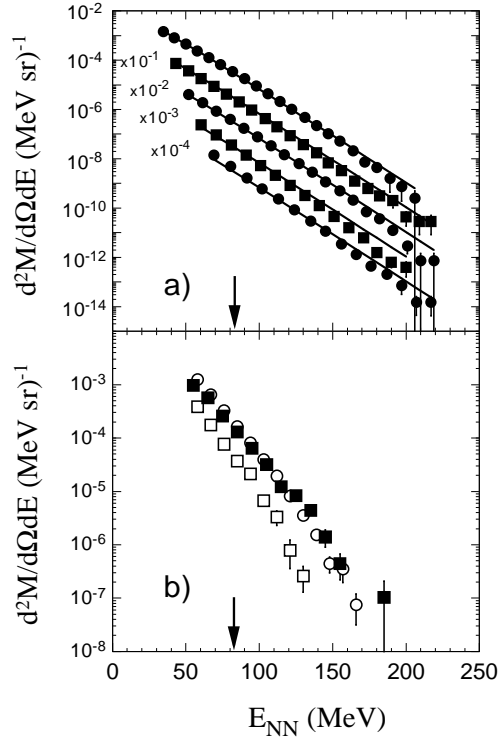,width=7.0cm}
\caption{
(a) Experimental inclusive proton multiplicity spectra in the NN reference 
frame at different polar angles ($87^o$, $110^o$, $113^o$, $126^o$, $140^o$).
(b) Experimental and BNV proton energy multiplicity spectrum in the 
NN reference frame for central collisions 
($<b> \approx 2.5$ fm, $98^o \leq  \theta_{NN} \leq 124^o$).
Experimental data (full squares), local Skyrme interaction (open squares) and 
momentum dependent interaction (open circles) are reported.
The arrows indicate the expected kinematical limit for first chance NN 
collisions.}
\end{figure}

Extremely energetic protons were already observed in heavy ion collisions 
(see \cite{CON00} and refs. therein), but those experiments did not allow
a conclusive answer on the production mechanism to be drawn.    
In order to explain the production of particles with energy or mass larger
than the energy available in NN collisions including the Fermi motion, 
several hypotheses, such as the presence of  high momentum tails in the
nucleon momentum distribution, the onset of multistep processes, cooperative 
effects, the presence of dynamical fluctuations, have been proposed. 

To gather more information on the energetic proton emission, we have performed 
simulations with a BNV code which is based on mean field and two body 
dynamics \cite{BON94}.
To explore the sensitivity to basic ingredients of the calculations
two different potentials were implemented in the BNV code: one using a 
local Skyrme interaction for the mean field (open squares of fig.~1b) 
and another using a Gale--Bertsh--Das Gupta momentum dependent interaction 
(GBD) (open circles of fig.~1b) \cite{GRE99}.  
Concerning the reaction dynamics, both calculations exhibit similar features
leading to the formation of a heavy residue in central collisions, whereas
binary collisions dominate at larger impact parameters. However,   
although momentum dependent effects are expected to be more important at
higher bombarding energy, it is interesting to understand how far 
their inclusion 
can affect the energetic proton production which is ruled by a delicate 
balance between the mean field and the nucleon-nucleon cross section. 
The experimental and calculated spectra in the NN frame are shown 
in fig.~1b for central collisions.
To achieve a satisfactory statistics for energetic proton production,
about 250 events have been simulated per impact parameter (using 200 test
particles per nucleon).  
The criteria for impact parameter selection and assignment 
will be discussed in detail in the following.
Our results point out remarkable differences: the spectrum calculated with the 
local Skyrme interaction strongly undershoots the data and
exhibits a lower inverse slope parameter than the experimental one, 
while the spectrum calculated 
with the momentum dependence is in better  agreement with the data concerning 
both the yield and the slope (at least up to $\approx 110$ MeV). 
Due to the less
attractive mean field, we observe in the GBD calculations a larger fraction 
of escaping particles. For the same reason, these particles can also be more 
energetic.
Moreover, the calculated yields should be slightly reduced since in this
kind of 
calculation only free nucleons are emitted while in the reaction also 
complex particles are emitted and observed experimentally \cite{WAD89}. 

From the comparison with the experimental data one can get
a deeper insight into the behaviour of nuclear matter at large density and 
temperature.
Therefore, with the aim of improving the overall understanding of the  
energetic proton emission and disentangling between the various
hypotheses for the production of the most energetic protons, we have 
investigated the impact parameter dependence.   
Indeed, the dependence of multiplicity on the number of nucleons participating 
in the reaction can provide information about a change in the production 
mechanism.
A stronger than linear increase of the multiplicity as a function of the 
number of participant nucleons has been observed, at much higher incident 
energy, in the deep subthreshold production of K$^+$ \cite{MIS94},
$\eta$ \cite{WOL98} and high transverse-mass $\pi^0$ \cite{MAR99}. 

At energy as low as 30A MeV, the fluctuations on global variables 
such as the charge particle multiplicity and transverse energy affect the 
determination of the
impact parameter especially for the most central collisions \cite{ADZ94}.
To solve the problem of the event selection over a wide range 
of impact parameters,  
we exploit the reaction mechanism and hard photon 
multiplicity information to determine the size of the interaction zone 
 \cite{NIF90,BAU86}. Indeed, the detection of heavy 
fragments from projectile-like fragments to evaporation residues and their
relative velocities allows to select classes of events with different
centrality. In particular, the most central collisions were selected requiring 
the presence of an evaporation residue with velocity close to the centre 
of mass velocity and charge higher than the projectile charge.
On the other hand, the detection of fragments originating from projectile
fragmentation and deep inelastic collisions  
was exploited to reach classes of
events spanning the range of impact parameters from peripheral to mid-central. 
The same procedure was also applied to some classes of events selected 
in terms of total charged particle multiplicity. 
Finally, to obtain a 
quantitative estimate of the impact parameter, the hard photon multiplicity 
has been calculated for the various classes of events. The number of 
participant nucleons $A_{part}(b)$ has been extracted from the hard photon 
($E_{\gamma} \geq 30 MeV$) multiplicity, according to the relation 
$M_{\gamma}(b) = P_{\gamma} \cdot N_{np}(b) \simeq 
P_{\gamma} \cdot 0.5 \cdot A_{part}(b)$ where $P_{\gamma}$ is the probability
of emitting a hard photon in a np collision deduced from inclusive data 
($P_{\gamma} (E_\gamma \geq 30 MeV) \approx 2.7 \cdot 10^{-5}$) and 
$N_{np}(b)$ is the number of first chance np collisions occurring in the 
overlap region. This relation has been  
satisfied by several experiments which show that the hard photon multiplicity
provides a snapshot of the participant region \cite{MIG93,PIA98,VAN96}.

In fig.~2 the average proton multiplicity is reported as
a function of the number of nucleons participating $A_{part}(b)$ in the 
collision for different energy bins in the NN reference frame 
($60 \div 80$ MeV $(M_p(60))$, 
$100 \div 120$ MeV $(M_p(100)$, 
$130 \div 150$ MeV $(M_p(130))$.  
The experimental proton multiplicity (full squares) displays the expected 
linear dependence on $A_{part}(b)$ \cite{ALB94} 
for energy close to the kinematical limit 
($60\leq E_p \leq 80$ MeV fig.~2a), while a stronger dependence is  
observed with increasing proton energy, in particular 
the multiplicity of extremely energetic protons $(M_p(130))$  
exhibits an almost quadratic increase with $A_{part}$ (fig.~2c).

The BNV calculations, filtered with the 
experimental apparatus, are also reported in fig.~2 (open circles). 
The $A_{part}(b)$ assignment relies on the hypothesis of a geometrical 
correlation between $b$ and $A_{part}(b)$ \cite{BAU86}. 
The calculations have been scaled by a factor 0.6 to allow a better
comparison with the data. This scaling is consistent with the yield reduction
needed to account for complex particle emission.
Within this assumption, a good agreement with the data is observed in fig.~2a 
and  fig.~2b,  
confirming that the energetic proton production is described with good accuracy 
up to $\approx$ 110 MeV. On the other hand, 
BNV calculations fail in the most energetic bin ($M_p(130)$, fig.~2c)
where the almost quadratic dependence on $A_{part}$ observed experimentally 
is not reproduced thus showing the onset of effects beyond the mean field and
two body collisions.  
It is interesting to notice that a non linear dependence is observed,
both experimentally and theoretically, also for $M_p (100)$.
The calculation can account for this behaviour due to the increasing importance 
of multistep two-body collisions in the production mechanism of protons with
energy higher than the kinematical limit \cite{MAR99}.
However, this mechanism seems not to be able to explain the almost quadratic
behaviour observed for $M_p (130)$.
Indeed, for extremely energetic protons, this multistep process is associated 
with larger time scale. Therefore the system can emit nucleons and 
rapidly evolves far from the initial geometrical overlap configuration. 
This can explain the weaker dependence on the impact parameter observed 
in the calculations (fig.~2c).

\begin{figure}[ht]
\vspace {0.0cm}
\psfig{figure=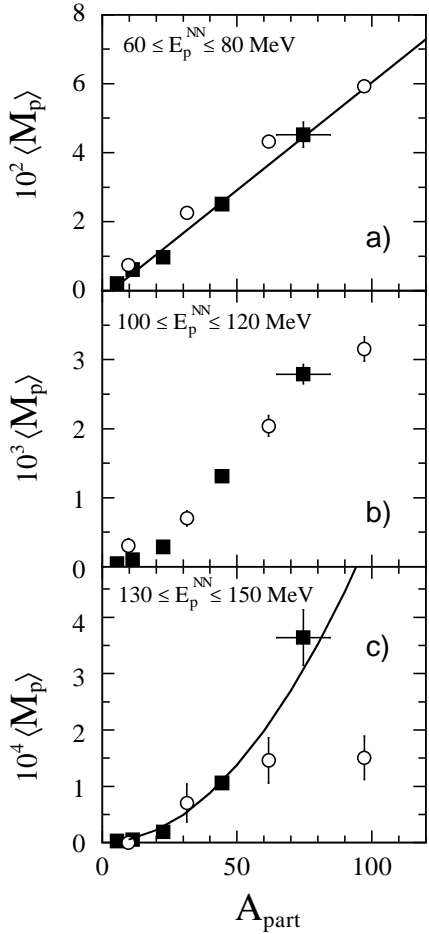,width=6.5cm}
\caption{Proton multiplicities as a function of the number of participant
nucleons (see text) for different energy bins. 
Experimental values (solid squares) and momentum dependent BNV calculations
(open circles) scaled by a factor 0.6 are reported. 
Only protons emitted in the angular range 
$75^o \leq \theta_{lab} \leq 138^o$ are considered.
To compare the trends a linear and a quadratic dependence are reported
(solid lines) in panels a) and c), respectively.}
\end{figure}

The observed behaviour of the multiplicity of
very energetic protons on the number of participant nucleons (fig. 2c)
puts constraints on the mechanism responsible for the production of 
extremely energetic protons. Dynamical 
fluctuations \cite{GER98} are not expected to lead 
to the $A_{part}$ quadratic behaviour observed experimentally.
Other effects, such as high momentum tails, are weakly dependent 
on density and, at the energy considered,
density variation from central to peripheral impact parameters are small 
\cite{BAL}.   
Cooperative effects, where more nucleons or clusters of nucleons
participate in the collision, seem very promising and should be investigated.

In summary, the energetic proton production has been investigated up to proton 
energy corresponding to about 20$\%$ of the total energy available in the
system.    
The energetic protons up to $\approx$ 110 MeV are 
emitted as a consequence of NN collisions in the first stage of the reaction
and their characteristics are well reproduced by BNV
calculations which include the momentum dependence in the effective potential.  
On the other hand, the BNV approach fails to explain the almost 
quadratic dependence on the number of participant nucleons of the yield of
very energetic protons ($E_p^{NN} \geq 130 MeV$). 
This behaviour calls for the introduction of  mechanisms beyond the mean field
and two body nucleon-nucleon collisions such as cooperative effects. 
These results shed some light on the emission of extremely
energetic protons and can improve the understanding of the mechanism responsible
for deep subthreshold particle production.

We thank Prof. M. Di Toro for the critical reading of the manuscript and the 
LNS staff for the high-quality beam and the support during the experiment. 


\begin{thebibliography}{99}

\bibitem {CAS90}
W. Cassing {\it et al.},
Phys. Rep. {\bf{188}}, 365 (1990).

\bibitem {GEL99}
C. Gelbke,
Prog. in Part. and Nucl. Phys. {\bf{42}}, 91 (1999).

\bibitem {FUC94}
H. Fucks and K. Mohring,
Rep. Prog. Phys. {\bf{57}}, 231 (1994).

\bibitem {WAD89}
R. Wada {\it et al.},
Phys. Rev. C {\bf{39}}, 497 (1989).

\bibitem {ALB94}
R. Alba {\it et al.},
Phys. Lett. B {\bf{322}}, 38 (1994).

\bibitem {SAP94}
P. Sapienza {\it et al.},
Phys. Rev. Lett. {\bf{73}}, 1769 (1994).

\bibitem {CON00}
R. Coniglione {\it et al.},
Phys. Lett. B {\bf{471}}, 339 (2000).

\bibitem {MAR99}
G. Martinez {\it et al.},
Phys. Rev. Lett. {\bf{83}}, 1538 (1999) and reference therein.

\bibitem {MIG92}
E. Migneco {\it et al.},
Nucl. Instr. and Meth., {\bf{A 314}}, 31 (1992). 

\bibitem {IOR93}
I. Iori {\it et al.},
Nucl. Inst. and Meth., {\bf{A 325}}, 458 (1993). 

\bibitem {ADZ93}
A. Del Zoppo {\it et al.},
Nucl. Inst. and Meth., {\bf{A 327}}, 363 (1993).

\bibitem {SAP98}
P. Sapienza {\it et al.},
Il Nuovo Cimento, {\bf{Vol. 111 A}}, 999 (1998).

\bibitem {BON94}
A. Bonasera {\it et al.},
Phys. Rep. {\bf{243}}, 1 (1994).

\bibitem {GRE99}
V. Greco {\it et al.},
Phys. Rev. C {\bf{59}}, 810 (1999).

\bibitem {MIS94}
D. Miskowiec {\it et al.},
Phys. Rev. Lett. {\bf{72}}, 3650 (1994).

\bibitem {WOL98}
A.R. Wolf {\it et al.},
Phys. Rev. Lett. {\bf{80}}, 5281 (1998).


\bibitem {ADZ94}
A. Del Zoppo {\it et al.},
Phys. Rev. C {\bf{50}}, 497 (1994).

\bibitem {NIF90}
H. Nifenecker and J.A.Pinston,
Annu. Rev. Nucl. Part. Sci. {\bf{40}}, 113 (1990).

\bibitem {BAU86}
W. Bauer {\it et al.},
Phys. Rev. C {\bf{34}}, 2127 (1986).

\bibitem {MIG93}
E. Migneco {\it et al.},
Phys. Lett. B {\bf{298}}, 46 (1993).

\bibitem {PIA98}
P. Piattelli {\it et al.},
Phys. Lett. B {\bf{442}}, 48 (1998).

\bibitem {VAN96}
J.H.G. van Pol {\it et al.},
Phys. Rev. Lett. {\bf{76}}, 1425 (1996).

\bibitem {GER98}
M. Germain {\it et al.},
Phys. Lett. B{\bf{437}}, 19 (1998).

\bibitem {BAL}
M. Baldo and U. Lombardo, private comunications.

\end{thebibliography}
\end{document}